\documentclass[conference]{IEEEtran}

\usepackage[english]{babel} 
\usepackage{graphicx}       
\usepackage{verbatim}       
\usepackage{amssymb}        
\usepackage[cmex10]{amsmath}        
\usepackage{amsxtra}        
\usepackage{amsthm} 
\usepackage{array}
\usepackage{url}
\usepackage[colorlinks]{hyperref}
\usepackage[caption=false,font=footnotesize]{subfig}

\usepackage{tikz}
\usetikzlibrary{calc,positioning}
\usetikzlibrary{spy}
\usepackage{dsfont}
\usepackage{bbm}
\usepackage{siunitx}
\usepackage{pgfplots}
\usepackage{bm}
\usepackage{booktabs}
\usepackage{cleveref}
\usepackage{mathtools}

\newcommand{\biAWGNC}{\text{biAWGNC}}

\newcommand{\HX}[1]{\mathbb{H}\left(#1\right)}

\newcommand{\I}[1]{\mathds{I}\left(#1\right)}

\newcommand{\abs}[1]{\left\lvert#1\right\rvert}

\newcommand{\cX}{\mathcal{X}}

\renewcommand{\d}[1]{\mathrm{d}#1}

\newcommand{\mH}{\boldsymbol{H}}

\newcommand{\va}{\boldsymbol{a}}

\newcommand{\rB}{\mathsf{B}}

\newcommand{\rL}{\mathsf{L}}

\newcommand{\rU}{\mathsf{U}}

\newcommand{\rX}{\mathsf{X}}
\newcommand{\rY}{\mathsf{Y}}
\newcommand{\rZ}{\mathsf{Z}}

\newcommand{\rvecB}{\bm{\mathsf{B}}}

\newcommand{\APP}{\text{APP}}

\newcommand{\ch}{\mathrm{ch}}
\newcommand{\Jf}[1]{\mathsf{J}\left(#1\right)}
\newcommand{\Jfi}[1]{\mathsf{J}^{-1}\left(#1\right)}

\theoremstyle{plain}

\pdfoutput=1

\usepackage{cite}
\pgfplotsset{compat=newest}
\newtheorem{remark}{Remark}
\newcommand\undermat[2]{%
  \makebox[0pt][l]{$\smash{\underbrace{\phantom{%
    \begin{matrix}#2\end{matrix}}}_{\text{$#1$}}}$}#2}

\begin{document}

\title{Protograph-Based LDPC Code Design\\for Bit-Metric Decoding}

\author{\IEEEauthorblockN{Fabian Steiner, Georg B\"ocherer}
\IEEEauthorblockA{Institute for Communications Engineering\\Technische Universit\"at M\"unchen, Germany\\
Email: \{fabian.steiner, georg.boecherer\}@tum.de}
\and
\IEEEauthorblockN{Gianluigi Liva}
\IEEEauthorblockA{Institute of Communication and Navigation of the\\ Deutsches
Zentrum f\"ur Luft- und Raumfahrt (DLR)\\ 82234 Wessling, Germany\\
Email: Gianluigi.Liva@dlr.de}
}
\IEEEoverridecommandlockouts

\maketitle

\begin{abstract}
A protograph-based low-density parity-check (LDPC) code design technique for bandwidth-efficient coded modulation is presented. The approach jointly optimizes the LDPC code node degrees and the mapping of the coded bits to the bit-interleaved coded modulation (BICM) bit-channels. For BICM with uniform input and for BICM with probabilistic shaping, binary-input symmetric-output surrogate channels are constructed  and used for code design. The constructed codes perform as good as multi-edge type codes of Zhang and Kschischang (2013). For 64-ASK with probabilistic shaping, a blocklength 64800 code is constructed that operates within $0.69$ dB of $\frac{1}{2}\log_2(1+\text{SNR})$ at a spectral efficiency of $4.2$ bits/channel use and a frame error rate of $10^{-3}$.
\end{abstract}
\section{Introduction}
\label{sec:intro}
Bit-interleaved coded modulation (BICM) combines high order modulation with binary error correcting codes \cite{zehavi_bicm}. This makes BICM
attractive for practical application and BICM is widely used in standards, e.g., in DVB-T2/S2/C2. At a BICM receiver, bit-metric decoding (BMD) is
used \cite[Sec. II]{martinez_mismatched}. Achievable rates for BMD were investigated for uniformly distributed inputs in \cite{martinez_mismatched} and for non-uniformly distributed bits in \cite{fabregas_bicm_shaping}. These results were generalized to non-uniformly distributed input symbols in \cite{boecherer_bmd} and
\cite{bocherer2014achievable}.

When designing low-density parity-check (LDPC) codes for BICM, the key challenge is to take into account the unequal error protection of the LDPC
coded bits and the BICM bit-channels that are different for different bit-levels. A first approach is to take an existing LDPC code and to 
optimize the mapping of the coded bits to the BICM bit-levels. This was done, e.g.,
in \cite{li2005,lei2008,hager2014optimized}. A more fundamental approach is to directly incorporate the different bit-channels in the code design.
This is done in \cite{zhang_MET_journal}, where the authors use multi-edge type (MET) codes \cite{richardson_MET} to parameterize the different
bit-channels. They then extend the extrinsic information transfer charts (EXIT) \cite{ten_brink_exit1} to multiple dimensions to design 
codes for quadrature amplitude modulation with 16 signal points (16-QAM). As 16-QAM can be constructed
as the Cartesian product of two four point amplitude-shift keying (4-ASK) constellations, two different bit-channels are apparent. For
constellations with more than two different bit-channels, the authors of \cite{zhang_MET_journal} observe long runtimes of their multidimensional EXIT approach. Therefore, they
suggest a high-order extension based on nesting, i.e., starting from $m=2$, they successively extend their codes from $m$ to $m + 1$ bit-levels by optimizing in each step only the additional
bit-level.

In this work, we follow \cite{zhang_MET_journal} and jointly optimize the code structure and the mapping of the coded bits to the BICM bit-levels. However, we propose a protograph-based design~\cite{thorpe_protograph} and use protograph EXIT (PEXIT) analysis \cite{liva_pexit} to determine the ensemble iterative convergence threshold by accounting for the different bit-channels associated with the
protograph variable nodes. The protograph ensemble is optimized with respect to the threshold by differential evolution. A design of 
protographs for coded modulation was first introduced in \cite{divsalar_error_floor_cm}, which relies on a variable degree matched
mapping (VDMM). More specifically, in \cite{divsalar_error_floor_cm} each bit-level is associated to a specific protograph variable 
node following the waterfilling approach (i.e., assigning the most protected coded bits to the bit-levels with highest bit-channel capacities). Recently, a 
protograph-based coded modulation scheme was introduced in \cite{marinoni2010proto} by performing a one-to-one mapping between the constellation 
symbols and the codeword symbols of a non-binary protograph LDPC code. This requires the constellation order to match the field order on which the 
LDPC code is constructed. 
To the best of our knowledge, none of the above-mentioned 
approaches leads to a joint  binary LDPC protograph ensemble and bit-mapping optimization.

For the PEXIT analysis, we represent each bit-channel by a surrogate. Our surrogates reflect both the BICM bit-channels and the input distribution. We optimize codes both for uniformly distributed inputs and for the probabilistic shaping strategy proposed in \cite{boecherer_bmd}. Our optimized codes show a finite length performance that is as good as the one presented by Zhang and
Kschischang \cite{zhang_MET_journal}. Moreover, our design approach can be applied to very large constellations. For 64-ASK with probabilistic shaping, our blocklength 64800 code operates within \SI{0.69}{\decibel} of 
$\frac{1}{2}\log_2(1+\text{SNR})$ at a spectral efficiency of \num{4.2} 
bits/channel use and a frame error rate of $10^{-3}$.

This paper is organized as follows. In Sec.~\ref{sec:bmd}, we review bit-metric decoding. We present our code design approach in Sec.~\ref{sec:surrogates}. In Sec.~\ref{sec:sim_results} and Sec.~\ref{sec:ps}, we discuss the performance of our codes for uniform and shaped inputs, respectively.
\section{Bit-Metric Decoding}
\label{sec:bmd}
Consider the discrete time additive white Gaussian noise (AWGN) channel
\begin{align}
\rY = \Delta \rX+\rZ\label{eq:awgnchannel}
\end{align}
where the noise term $\rZ$ is zero mean, unit variance Gaussian, and where the input $\rX$ is distributed on the normalized $2^m$-ASK constellation
\begin{align}
\cX=\{\pm 1,\pm 2,\dotsc,\pm (2^m-1)\}.
\end{align}
The constellation spacing $\Delta$ controls the average power $\mathsf{E}[|\Delta 
\rX|^2]$, where $\mathsf{E}[\cdot]$ denotes expectation. The 
signal-to-noise ratio (SNR) is $\text{SNR}=\mathsf{E}[\abs{\Delta 
\rX}^2]/1$. Each signal point 
$x\in \cX$ is labeled by $m$ bits $\rvecB = (\rB_1, \rB_2,\dotsc, \rB_m)$. The $\rB_i$ represent the bit-levels. Throughout 
this work, we label by the binary reflected Gray code (BRGC) \cite{gray1953pulse}, e.g., see Fig.~\ref{fig:8ASK_constellation_labeling}. Let $p_{\rY|\rvecB}$ be the memoryless channel with random
input $\rvecB$ and output $\rY$. The bit-metric of bit-level $i$ is the $L$-value
\begin{align}
L_i=\log\frac{P_{\rB_i|\rY}(0|y)}{P_{\rB_i|\rY}(1|y)}\label{eq:lvalue}
\end{align}
where $P_{\rB_i|\rY}$ can be calculated from the joint distribution
\begin{align}
p_{\rB_i\rY}(b_i,y)=\sum_{\va\in\{0,1\}^m\colon a_i=b_i}p_{\rY|\rvecB}(y|\va)P_{\rvecB}(\va).\label{eq:pbiy}
\end{align}
A bit-metric decoder uses the $L$-values $L_1, L_2,\dotsc, L_m$ to estimate the transmitted data. For the decoder, the channel appears as $m$ 
parallel bit-channels, see Fig.~\ref{fig:bitchannels}.
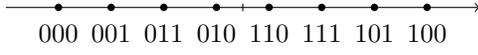
\begin{figure}
\centering
\begin{tikzpicture}[scale=.7]
 \draw[->] (-1,0) -- (8,0);
 \foreach \a/\b in {$000$/0, $001$/1, $011$/2, $010$/3, $110$/4, $111$/5, $101$/6, $100$/7} {
  \fill[black] (\b,0) circle (2pt);
  \node at (\b,-.5) {\a};
 }
 \draw (3.5,-2pt) -- (3.5,2pt);
\end{tikzpicture}
\caption{8-ASK constellation with BRGC \cite{gray1953pulse} labeling.}
\label{fig:8ASK_constellation_labeling}
\end{figure}
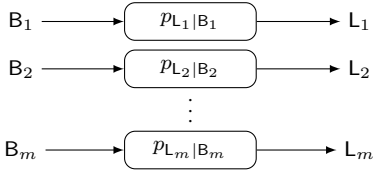
\begin{figure}
\centering
\footnotesize
\begin{tikzpicture}[scale=.9]
\node(B1)at (-5,5){$\rB_1$};

\path (B1) --++ (2.5,0) node (ch1)  [rectangle,draw,minimum width=1.75cm,minimum height=0.5cm,rounded corners] {$p_{\rL_1|\rB_1}$} --++(2.5,0) node 
(L1){$\rL_1$};

\path (B1) --++(0,-0.7) node (B2){$\rB_2$} --++ (2.5,0) node (ch2) [rectangle,draw,minimum width=1.75cm,minimum height=0.5cm,rounded corners,minimum height=0.5cm] {$p_{\rL_2|\rB_2}$} --++(2.5,0) node (L2){$\rL_2$};

\path (B2) --++(0,-1.2) node (Bm){$\rB_m$} --++ (2.5,0) node (chm) [rectangle,draw,minimum width=1.75cm,minimum height=0.5cm,rounded corners] {$p_{\rL_m|\rB_m}$} --++(2.5,0) node (Lm){$\rL_m$};
\path (ch2) --++(0,-0.5) node{$\vdots$};

\draw[-latex] (B1) -- (ch1);\draw[-latex] (ch1) -- (L1);
\draw[-latex] (B2) -- (ch2);\draw[-latex] (ch2) -- (L2);
\draw[-latex] (Bm) -- (chm);\draw[-latex] (chm) -- (Lm);

\end{tikzpicture}
\caption{$m$ parallel bit-channels that are different for different bit-levels.}
\label{fig:bitchannels}
\end{figure}
Bit-metric decoding can achieve the rate \cite[Theorem~1]{bocherer2014achievable}
\begin{align}
R_{\textnormal{BMD}}=\HX{\rvecB}-\sum_{i=1}^m \HX{\rB_i|\rL_i}\label{eq:bmdrate}
\end{align}
where $\HX{\cdot}$ denotes entropy.
\begin{remark}
If the bits $\rB_1,\rB_2,\dots,\rB_m$ are independent, then \eqref{eq:bmdrate} can be written as \cite{martinez_mismatched,fabregas_bicm_shaping}
\begin{align}
R_{\textnormal{BMD}}=\sum_{i=1}^m\I{\rB_i;\rL_i}
\end{align}
where $\I{\cdot;\cdot}$ denotes mutual information.
\end{remark}
\begin{remark}
If the bits $\rB_1,\rB_2,\dots,\rB_m$ are independent and uniformly distributed, then the $L$-values \eqref{eq:lvalue} can be written as
\begin{align}
L_i=\log\frac{p_{\rY|\rB_i}(y|0)}{p_{\rY|\rB_i}(y|1)}.
\end{align}
\end{remark}
We use $L$-values as defined in \eqref{eq:lvalue} and the achievable rate as defined in \eqref{eq:bmdrate}, 
since we will use non-uniformly distributed inputs in Sec.~\ref{sec:ps}.

\section{LDPC Code Design}
\label{sec:surrogates}
LDPC code ensembles are usually characterized by the degree profiles of the variable and check nodes of the 
Tanner graph representation of the sparse binary parity-check matrix $\mH\in\mathds{F}_2^{(n-k)\times n}$~\cite[Section 3.3]{mct}. For instance, 
$\lambda(x)=\sum_{d=1}^{d_v} \lambda_d x^{d-1}$ and $\rho(x)=\sum_{d=1}^{d_c} \rho_d x^{d-1}$ are the edge-perspective variable and check node degree polynomials with maximum degree $d_v$ and $d_c$, respectively. However, the degree profiles do not characterize the mapping of variable 
nodes to different bit-channels. In the following, we use protographs to incorporate the bit-mapping in our threshold analysis.  

\subsection{Protograph-Based LDPC Codes}
\label{sec:protograph}

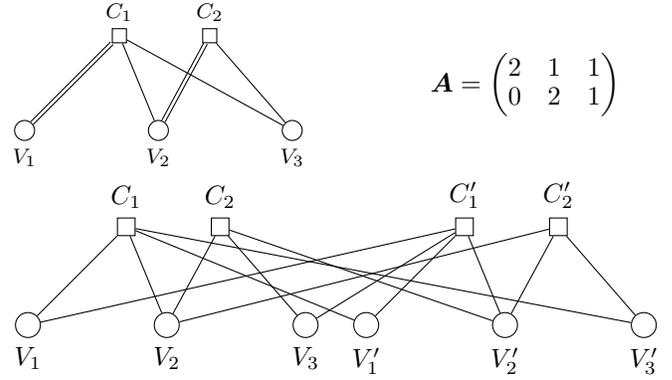
\begin{figure}
\begin{minipage}{.3\textwidth}
\footnotesize
\begin{tikzpicture}
 \node[rectangle,draw,thin,label={above:$C_1$}] (C1) {};
 \node[rectangle,draw,thin,label={above:$C_2$}] (C2) [right=of C1] {};
 
 \node[circle,draw,thin,label={below:$V_1$}] (V1) [below left=1.5cm of C1] {};
 \node[circle,draw,thin,label={below:$V_2$}] (V2) [right=1.5cm of V1] {};
 \node[circle,draw,thin,label={below:$V_3$}] (V3) [right=1.5cm of V2] {};

 \draw[double distance=1pt] (V1) -- (C1);
 \draw (V2) -- (C1);
 \draw[double distance=1pt] (V2) -- (C2);
 \draw (V3) -- (C1);
 \draw (V3) -- (C2);
\end{tikzpicture}
\end{minipage}
\begin{minipage}[b]{.1\textwidth}
 \begin{equation*}
 \boldsymbol{A} = \begin{pmatrix}
        2 & 1 & 1\\
        0 & 2 & 1
       \end{pmatrix}\label{eq:protograph_basematrix}
\end{equation*}
\end{minipage}

\begin{tikzpicture}
 \node[rectangle,draw,thin,label={above:$C_1$}] (C1) {};
 \node[rectangle,draw,thin,label={above:$C_2$}] (C2) [right=of C1] {};
 
 \node[circle,draw,thin,label={below:$V_1$}] (V1) [below left=1.5cm of C1] {};
 \node[circle,draw,thin,label={below:$V_2$}] (V2) [right=1.5cm of V1] {};
 \node[circle,draw,thin,label={below:$V_3$}] (V3) [right=1.5cm of V2] {};

\node[rectangle,draw,thin,label={above:$C_1'$}] (C1p) [right=3	cm of C2] {};
 \node[rectangle,draw,thin,label={above:$C_2'$}] (C2p) [right=of C1p] {};
 
 \node[circle,draw,thin,label={below:$V_1'$}] (V1p) [below left=1.5cm of C1p] {};
 \node[circle,draw,thin,label={below:$V_2'$}] (V2p) [right=1.5cm of V1p] {};
 \node[circle,draw,thin,label={below:$V_3'$}] (V3p) [right=1.5cm of V2p] {};

 \draw (V1) -- (C1);
 \draw (V2) -- (C1);
 \draw (V2) -- (C2);
 \draw (V3) -- (C2);

 \draw (V1p) -- (C1p);
 \draw (V2p) -- (C1p);
 \draw (V2p) -- (C2p);
 \draw (V3p) -- (C2p);

\draw (V1) -- (C1p);
\draw (V2) -- (C2p);
\draw (V1p) -- (C1);
\draw (V2p) -- (C2);

\draw (V3p) -- (C1);
\draw (V3) -- (C1p);

\end{tikzpicture}
\caption{Above, the tanner graph of basematrix $\boldsymbol{A}$ with $e=2$ and $f=3$ is displayed. Below, an example lifting with $Q=2$ instances of the protograph is shown.}
\label{fig:protograph_base}
\end{figure}

Starting from a small bipartite graph represented via its basematrix $\boldsymbol{A}=[b_{lk}]$ of size $e\times f$, one applies a copy-and-permute operation (also known as lifting) to create $Q$ instances of the small graph and then permutes the edges so that the local edge connectivity remains the same. The $Q$ replicas of variable 
node $V_k, k\in\{1,\ldots,f\}$ must be connected only to replicas of the neighbors of $V_k$ while maintaining the original degrees for that specific edge. The resulting bipartite graph representing the
final parity-check matrix $\mH$ possesses $n = Q \cdot f$ variable nodes and $n-k = Q\cdot e$ check nodes. Parallel edges are allowed, but must be resolved during the copy-and-permute
procedure. An example protograph with the corresponding basematrix and an example lifting for $Q=2$ are shown in Figure~\ref{fig:protograph_base}.

\subsection{Surrogate Channels}
\label{sec:matching}
\begin{figure*}
\centering
\newcounter{myeq1}
\setcounter{myeq1}{\value{equation}}
\setcounter{equation}{5}
\setcounter{equation}{\value{myeq1}}
 \small
 \begin{align}
 I_{E_{V_k\to C_l}} &= \Jf{\sqrt{\sum_{\substack{l'=1\\l'\neq l}}^{e} b_{l'k} \cdot \Jfi{I_{E_{C_{l'}\to V_k}}}^2 + (b_{lk}-1) \cdot \Jfi{I_{E_{C_l\to V_k}}}^2 + \sigma^2_{\ch_k}}}\label{eq:pexit_v2c}\\
 I_{E_{C_l\to V_k}} &= 1-\Jf{\sqrt{\sum_{\substack{k'=1\\k'\neq k}}^{f} b_{lk'} \Jfi{1-I_{E_{V_{k'}\to C_{l}}}}^2 + (b_{lk}-1) \cdot \Jfi{1-I_{E_{V_k\to C_l}}}^2}}\label{eq:pexit_c2v}\\
 I_{\APP_k} &= \Jf{\sqrt{\sum_{l'=1}^{f} b_{l'k} \cdot \Jfi{I_{E_{C_{l'}\to V_k}}}^2 + \sigma^2_{\ch_k}}}.\label{eq:pexit_iapp}
 \end{align}
 \hrulefill
\end{figure*}
Optimizing LDPC codes directly for the $m$ bit-channels in Fig.~\ref{fig:bitchannels} is difficult, as they are usually not output-symmetric and do not possess uniform input in the case of probabilistic shaping. We instead optimize for a surrogate channel. The optimized code works well on the original channel if the surrogate channel preserves some characteristics of the original channel and if the code is universal in the sense that its performance depends mainly on these characteristics. We propose to use the \emph{rate backoff} as the key characteristic, i.e., if $R^*$ is the achievable rate of the original channel and $R$ is the actual transmission rate of the code, we assume that the decoding error probability depends on the rate backoff
\begin{align}
R^*-R.
\end{align}
We design codes for a surrogate channel with the same rate backoff. Let $\tilde{R}$ be the transmission rate and $\tilde{R}^*$ the achievable rate of the surrogate channel; we require
\begin{align}
\tilde{R}^*-\tilde{R}\overset{!}{=}R^*-R.
\end{align}
Suppose our code rate is $c$ so that $(1-c)m$ bits per channel use consist of redundancy bits on average. Thus, the transmission rate is
\begin{align}
\HX{\rvecB}-(1-c)m.\label{eq:transmissionrate}
\end{align}
For instance, for uniformly distributed inputs, the transmission rate is $R=cm$. By \eqref{eq:bmdrate} and \eqref{eq:transmissionrate}, the rate backoff is
\begin{align}
R^*-R=(1-c)m-\sum_{i=1}^m\HX{\rB_i|\rL_i}.
\end{align}
Note that the term $(1-c)m$ does not depend on the channel. Thus, we operate the surrogate channel at the same rate backoff if
\begin{align}
\sum_{i=1}^m\HX{\rB_i|\rL_i}\overset{!}{=}\sum_{i=1}^m\HX{\tilde{\rB}_i|\tilde{
\rY}_i}.\label{eq:ratebackoffcond}
\end{align}
Since we want to account for the different bit-channels, we restrict our surrogate channel further by requiring equality for the individual conditional entropies in \eqref{eq:ratebackoffcond} for each bit-level. We use biAWGN channels with uniform input as surrogate channel, i.e., we have
\begin{align}
\tilde{\rY}_i=x_{\tilde{\rB}_i}+\rZ_i
\end{align}
where the $\tilde{\rB}_i$ are binary and uniformly distributed, $x_0=1$, $x_1=-1$, and where $\rZ_i$ is zero mean Gaussian with variance 
$\sigma_{B_i}^2$ where
\begin{align}
\sigma_{B_i}^2\colon\quad\HX{\rB_i|\rL_i}\overset{!}{=}\HX{\tilde{\rB}
_i|\tilde{\rY}_i}.\label{eq:surrogate}
\end{align}

\subsection{PEXIT Analysis}
\label{sec:pexit}

The analysis of the asymptotic decoding threshold can be done by PEXIT analysis~\cite{liva_pexit}. For EXIT
chart analysis, the $\mathsf{J}$ function is widely used. The value $\Jf{\sigma_\ch}$ hereby represents the capacity of the \biAWGNC{} with variance $\sigma^2$, where $\sigma_{\ch}^2=\frac{4}{\sigma^2}$:
\begin{equation}
 \Jf{\sigma_{\ch}} = 1 - \int\limits_{-\infty}^\infty \frac{1}{\sqrt{2\pi\sigma^2_{\ch}}} \mathrm{e}^{-\frac{\left(\ell-\sigma^2_{\ch}/2\right)^2}{2\sigma^2_{\ch}}}\log_2\left(1+\mathrm{e}^{-\ell}\right)\d{\ell}.
\end{equation}
PEXIT analysis requires tracking the extrinsic mutual information values $I_E$ per $(V_k, C_l)$ pair to accommodate for the edge connections imposed by the protograph basematrix and the (possibly different) reliability of the coded bits $V_k$. The PEXIT algorithm is summarized as follows:
\begin{enumerate}
\item Initialize each $\sigma_{\ch_k}, k\in\{1,\ldots,f\}$ with a corresponding 
$\sigma_{B_i}, i\in\{1,\ldots,m\}$ \eqref{eq:surrogate} originating from the 
biAWGNC surrogates. For all tuples 
$(V_k,C_l), k\in\{1,\ldots,f\}, l\in\{1,\ldots,e\}$, set the a priori mutual 
information values $I_{E_{C_l\to V_k}} = 0$.
\item Perform variable to check node update \eqref{eq:pexit_v2c} for all tuples 
$(V_k,C_l)$.
\item Perform check to variable node update \eqref{eq:pexit_c2v} for all tuples 
$(V_k,C_l)$.
\item Calculate $I_{\APP_k}, k\in\{1,\ldots,f\}$ \eqref{eq:pexit_iapp}.
\item Stop if all $I_{\APP_k}=1$, otherwise go to 2).
\end{enumerate}

\section{Code Design for Uniform Inputs}
\label{sec:sim_results}
\subsection{Optimization Procedure for Protograph Ensembles}
\label{sec:optimization_procedure}
\begin{figure*}
 \centering
 \renewcommand{\arraystretch}{1.0}
 \footnotesize
 \begin{tabular}{p{3cm}p{8cm}p{2cm}p{2cm}}
  Constellation (code rate) & Basematrix & PEXIT Decoding Threshold & Gap to 
$R_\text{BMD}~\eqref{eq:bmdrate}$\\
  \toprule
  4-ASK (1/2) & $\boldsymbol{A} = \left(\begin{array}{llllll}
2 & 1 & 1 & 2 & 1 & 4\\
1 & 1 & 1 & 2 & 2 & 5\\
\undermat{B_2}{1 & 0 & 0} & \undermat{B_1}{1 & 0 & 6}
\end{array}\right)$\vspace{.5cm} & \SI{5.57}{\decibel} & \SI{0.28}{\decibel}\\
  4-ASK (3/4) & $\boldsymbol{A} = \left(\begin{array}{llllllll}
        1 & 1 & 1 & 1 & 6 & 6 & 1 & 1\\
        \undermat{B_2}{1 & 1 & 2 & 2} & \undermat{B_1}{6 & 6 & 2 & 2}
\end{array}\right)$\vspace{.5cm} &  \SI{9.57}{\decibel}& \SI{0.26}{\decibel}\\
\midrule
  64-ASK (5/6) & $\boldsymbol{A} = \left(\begin{array}{llllllllllll}
        2 & 2 & 2 & 1 & 2 & 2 & 6 & 2 & 2 & 0 & 6 & 6\\
        \undermat{B_2}{1 & 1} & \undermat{B_3}{1 & 2} & \undermat{B_4}{1 & 1} & \undermat{B_5}{6 & 1} & \undermat{B_6}{0 & 2} & \undermat{B_1}{6 & 6}
       \end{array}\right)$\vspace{.5cm} & \SI{25.52}{\decibel} & \SI{0.29}{\decibel}\\
 \end{tabular}
 \caption{Optimized protographs and design parameters.}
 \label{tab:results}
\end{figure*}
\begin{figure*}
\begin{minipage}{0.98\columnwidth} 
\footnotesize
%
%
\begin{tikzpicture}[scale=.9]

\begin{axis}[%
scale only axis,
xmin=5.1,
xmax=6.3,
xlabel={SNR [dB]},
xmajorgrids,
xminorgrids,
ymode=log,
ymin=1e-06,
ymax=1,
yminorticks=true,
ylabel={BER/FER},
ymajorgrids,
yminorgrids,
legend style={at={(axis cs:5.3,2e-6)},anchor=south west,legend cell align=left,align=left,draw=white!15!black}
]
\addplot [color=red,solid,line width=1.0pt,mark=+]
  table[row sep=crcr]{%
5.4	0.137416604892513\\
5.5	0.122621695082777\\
5.6	0.118477678939132\\
5.7	0.105618359278478\\
5.8	0.0478492298822173\\
5.9	0.0170130595136773\\
6	0.00459110658100651\\
6.1	0.000390074325710522\\
6.2	2.54981470022915e-05\\
6.3	0\\
6.4	0\\
};
\addlegendentry{MET code \cite{zhang_MET_journal} BER};

\addplot [color=red,dashed,line width=1.0pt,mark=+]
  table[row sep=crcr]{%
5.4	1\\
5.5	0.958333333333333\\
5.6	0.958333333333333\\
5.7	0.916666666666667\\
5.8	0.479166666666667\\
5.9	0.185185185185185\\
6	0.04375\\
6.1	0.00437445319335083\\
6.2	0.00025507601265177\\
6.3	0\\
6.4	0\\
};
\addlegendentry{MET code \cite{zhang_MET_journal} FER};

\addplot [color=blue,solid,line width=1.0pt,mark=o,mark options={solid}]
  table[row sep=crcr]{%
5.4	0.178343621399177\\
5.5	0.162289094650206\\
5.6	0.123533950617284\\
5.7	0.0579115226337448\\
5.8	0.0137288523090992\\
5.9	0.00218504489337823\\
6	0.000450824850191322\\
6.1	4.36436436436436e-05\\
6.2	3.44416891093157e-07\\
};
\addlegendentry{Optimized Protograph BER};

\addplot [color=blue,dashed,line width=1.0pt,mark=o,mark options={solid}]
  table[row sep=crcr]{%
5.4	1\\
5.5	1\\
5.6	1\\
5.7	0.555555555555556\\
5.8	0.194444444444444\\
5.9	0.0378787878787879\\
6	0.00730994152046784\\
6.1	0.000900900900900901\\
6.2	0.000169986401087913\\
};
\addlegendentry{Optimized Protograph FER};

\addplot [color=black,solid]
  table[row sep=crcr]{%
5.2805	1e-06\\
5.2805	0.01\\
};
\addlegendentry{16-ASK uniform BMD};

\end{axis}
\end{tikzpicture}%
 \caption{16-ASK: Performance of a rate 1/2 optimized protograph code compared to the rate 1/2 MET code in \cite{zhang_MET_journal}.}
 \label{fig:zhang12}
\end{minipage}
\hfill
\begin{minipage}{0.98\columnwidth} 
 \footnotesize
%
%
\begin{tikzpicture}[scale=.9]

\begin{axis}[%
scale only axis,
xmin=9.2,
xmax=10,
xlabel={SNR [dB]},
xmajorgrids,
xminorgrids,
ymode=log,
ymin=1e-06,
ymax=1,
yminorticks=true,
ylabel={BER/FER},
ymajorgrids,
yminorgrids,
legend style={at={(axis cs:9.34,2e-6)},anchor=south west,legend cell align=left,align=left,draw=white!15!black}
]
\addplot [color=blue,solid,line width=1.0pt,mark=+]
  table[row sep=crcr]{%
9.3	0.0660785032019972\\
9.4	0.0632292684250516\\
9.5	0.0496648757190926\\
9.6	0.0284872598502117\\
9.7	0.00754800528896914\\
9.8	0.00221117201538888\\
9.9	0.000155186803955793\\
10	8.48837301518254e-06\\
};
\addlegendentry{MET code \cite{zhang_MET_journal} BER};

\addplot [color=blue,dashed,line width=1.0pt,mark=+]
  table[row sep=crcr]{%
9.3	1\\
9.4	1\\
9.5	0.833333333333333\\
9.6	0.520833333333333\\
9.7	0.159090909090909\\
9.8	0.0462962962962963\\
9.9	0.00345065562456867\\
10	0.000363504180298073\\
};
\addlegendentry{MET code \cite{zhang_MET_journal} FER};

\addplot [color=red,solid,line width=1.0pt,mark=o,mark options={solid}]
  table[row sep=crcr]{%
9.3	0.0672118124041201\\
9.4	0.0599454854262547\\
9.5	0.0501212195923734\\
9.6	0.029268481992841\\
9.7	0.00915447622178392\\
9.8	0.0018607169381742\\
9.9	0.000150110492034352\\
10	7.55752937078782e-06\\
10.1	2.40036681352145e-06\\
};
\addlegendentry{Optimized Protograph BER};

\addplot [color=red,dashed,line width=1.0pt,mark=o,mark options={solid}]
  table[row sep=crcr]{%
9.3	1\\
9.4	1\\
9.5	0.916666666666667\\
9.6	0.638888888888889\\
9.7	0.2\\
9.8	0.0462962962962963\\
9.9	0.00339443312966735\\
10	0.000434593654932638\\
10.1	0.000179985601151908\\
};
\addlegendentry{Optimized Protograph FER};

\addplot [color=black,solid]
  table[row sep=crcr]{%
9.308	1e-06\\
9.308	0.01\\
};
\addlegendentry{16-ASK uniform BMD};

\end{axis}
\end{tikzpicture}%
 \caption{16-ASK: Performance of a rate 3/4 optimized protograph code compared to the rate 3/4 MET code in \cite{zhang_MET_journal}.}
 \label{fig:zhang34}
\end{minipage}
\end{figure*}
We impose restrictions on the dimension of the basematrix $\boldsymbol{A} \in \{0,1,\ldots,S\}^{e\times f}$. The parameters $e$ and $f$ determine how many different variable degrees per bit-channel can occur and are chosen to represent the desired code rate $c=\frac{f-e}{f}$. The maximum number of parallel edges $S$ is crucial for both the performance of the code and the success of the optimization procedure. The product $e\cdot S$ describes the maximum 
variable node degree in the final code and thereby determines the size of the optimization search space. The optimization of the basematrices is performed by differential evolution (DE), see \cite{richardson_design_irregular}.

\subsection{Numerical Results}
\label{subsec:sim_results}

In order to compare our code design approach to the setting of Zhang and 
Kschischang~\cite{zhang_MET_journal}, we design codes of rates 1/2 and 
3/4 for 4-ASK constellations with uniform inputs. The 
optimized protographs are depicted in Table~\ref{tab:results}. We discuss the rate 1/2 code in more detail. The coded bits are transmitted over two different bit-channels. For each bit-channel, we allow 3 (possibly different) variable degrees so that always 3 variable nodes in the protograph are assigned to the same surrogate channel: $\sigma_{\ch_1}^2=\sigma_{\ch_2}^2 = 
\sigma_{\ch_3}^2= 4/\sigma_{B_2}^2$, $\sigma_{\ch_4}^2=\sigma_{\ch_5}^2 = \sigma_{\ch_6}^2= 4/\sigma_{B_3}^2$.
Once the optimized basematrices have been found, we construct quasi-cyclic 
parity-check matrices with blocklengths $n=16200$. We simulate the constructed codes using 100 decoding  
iterations. The bit error rates (BER) and frame error rates (FER) in Fig.~\ref{fig:zhang12} and \ref{fig:zhang34} show that the finite length performance of our codes is equal to or slightly 
better than the codes in \cite{zhang_MET_journal}. The decoding thresholds given in Table~\ref{tab:results} 
are obtained by PEXIT analysis for the surrogate channels.

\section{Code Design for Shaped Inputs}
\label{sec:ps}
\newcommand{\kd}{k_{\!\text{d}}}
\newcommand{\nc}{n_{\!\text{c}}}

We briefly outline how our design procedure can be used to design LDPC 
codes for the probabilistic shaping scheme proposed in \cite{boecherer_bmd}.

\subsection{Probabilistic Shaping}
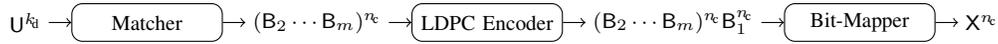
\begin{figure*}
\footnotesize
\centering
\begin{tikzpicture}
\draw (0,0) node (data) {$\rU^{\kd}$};
\draw node[draw,right=0.3cm of data, rectangle,rounded corners,minimum 
width=2cm,minimum height=.5cm] (matcher) {Matcher};
\draw node[right=0.3cm of matcher] (ms) {$(\rB_2\dotsb \rB_m)^{\nc}$};
\draw node[draw,right=0.3cm of ms, rectangle,rounded corners,minimum 
width=2cm,minimum 
height=0.5cm] (enc) {LDPC 
Encoder};
\draw node[right=0.3cm of enc] (encs) {$(\rB_2\dotsb 
\rB_m)^{\nc}\rB_{1}^{\nc}$};
\draw node[draw,right=0.3cm of encs,rectangle,rounded corners,minimum 
width=2cm,minimum 
height=0.5cm] (bm) 
{Bit-Mapper};
\draw node[right=0.3cm of bm] (x) {$\rX^{\nc}$};
\draw[->] (data) -- (matcher);
\draw[->] (matcher) -- (ms);
\draw[->] (ms) -- (enc);
\draw[->] (enc) -- (encs);
\draw[->] (encs) -- (bm);
\draw[->] (bm) -- (x);
\end{tikzpicture}
\caption{Probabilistic shaping as proposed in \cite{boecherer_bmd}. Independent  
uniformly distributed data bits $\rU^{\kd}=\rU_1\rU_2\dotsb \rU_{\kd}$ are 
matched to $\nc$ strings $(\rB_2\rB_3\dotsb \rB_m)_i$, $i=1,2,\dotsc,\nc$ that 
are distributed according to $P_{\rB_2\dotsb \rB_m}$. The systematic LDPC encoder appends $\nc$ check bits 
$\rB_1^{\nc}=\rB_{11}\rB_{12}\dotsb \rB_{1\nc}$. The bit mapper maps this 
bit stream to signal points according to the BRGC labeling. The overall rate is 
$\kd/{\nc}$ [bits/channel use] and the rate of the LDPC code is 
$(m-1)\nc/(m\nc)=(m-1)/m$. At the receiver, a bit-metric decoder calculates an 
estimate of $(\rB_2\dotsb \rB_m)^{\nc}\rB_1^{\nc}$. A data estimate 
$\hat{\rU}^{\kd}=\hat{\rU}_1\hat{\rU}_2\dotsb \hat{\rU}_{\kd}$ is obtained 
by passing the estimate of $(\rB_2\dotsb \rB_m)^{\nc}$ through a dematcher. For the matcher and 
the dematcher, we use the implementation \url{http://www.lnt.ei.tum.de/fileadmin/staff/gu92vox/matcher.zip}.}
\label{fig:ps}
\end{figure*}
Fig.~\ref{fig:ps} illustrates the shaping scheme.  
The key observation is that the capacity-achieving distribution of 
ASK constellations in Gaussian noise is symmetric around the origin. 
Consequently, the capacity-achieving distribution induces a distribution 
$P_{\rB_1\rB_2\dotsb \rB_m}$ on the BRGC labeling with the following properties:
\begin{itemize}
\item bit-level $\rB_1$ is uniformly distributed (bit-level $\rB_1$ decides on 
the sign of the transmitted constellation point, see 
Fig.~\ref{fig:8ASK_constellation_labeling}).
\item bit-levels $(\rB_2\dotsb \rB_m)$ and bit-level $\rB_1$ are independent. bit-levels $B_2,\dotsc,B_m$ are correlated.
\end{itemize}
This suggests to mimic the capacity-achieving distribution in the following way: 
first, generate bit-levels $\rB_2\dotsb \rB_m$ according to $P_{\rB_2\dotsb 
\rB_m}$, e.g., by using a distribution matcher (see 
Fig.~\ref{fig:ps}). Then encode by a systematic encoder of rate $(m-1)/m$ that copies the bits $\rB_2\dotsb \rB_m$ to its output and 
leaves their distribution un-changed. The encoder appends check bits $\rB_1$ that 
are approximately uniformly distributed because each check bit is a modulo two 
sum of many information bits \cite[Sec.~7.1]{boecherer_thesis}. The signal point $x_{\rB_1\rB_2\dotsb \rB_m}$ 
selected by the bit-mapper then has approximately the capacity-achieving 
distribution.

\subsection{Surrogate Channels}
As discussed before, the channel appears to the receiver as a set of $m$ 
parallel binary input channels, see Fig.~\ref{fig:bitchannels}. The non-uniform distribution $P_{\rB_2\dotsb 
\rB_m}$ influences each channel law $p_{\rL_i|\rB_i}$ via \eqref{eq:pbiy}. 
Consequently, it also influences the conditional entropies 
$\HX{\rB_i|\rL_i}$ in our surrogate channels \eqref{eq:surrogate} that we use for code optimization.

\subsection{Numerical Results}

\begin{figure}
 \footnotesize
 \input{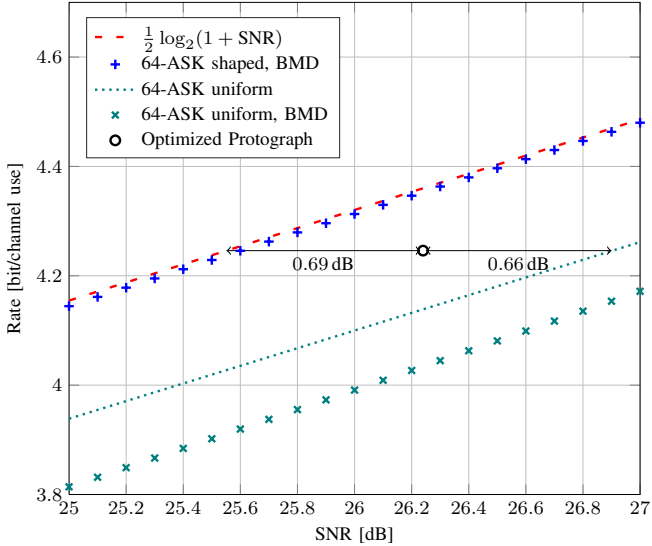}
 \caption{64-ASK: The operating point of an optimized protograph code of blocklength $n=64800$ and FER$=10^{-3}$. For comparison, AWGN capacity and power-rate curves are displayed.}
 \label{fig:ASK64}
\end{figure}

We optimize codes for 64-ASK with probabilistic shaping. To obtain a close-to-optimal input distribution, we use the approach of \cite{kschischang_pasupathy_maxwell},\cite[Sec.~3]{bocherer2014achievable} with a sampled Gaussian (Maxwell-Boltzmann) distribution. We maximize the resulting BMD rate \eqref{eq:bmdrate} over the constellation spacing $\Delta$ in \eqref{eq:awgnchannel}. The results are shown in Fig.~\ref{fig:ASK64}. Our optimized protograph code of blocklength $n=64800$ shows a gap of \SI{0.69}{\decibel} to AWGN capacity at \num{4.2} bits/channel use for FER$=10^{-3}$. Our operating point is \SI{0.66}{\decibel} more energy efficient than ideal uniform 64-ASK and \SI{1.17}{\decibel} more energy efficient than ideal uniform 64-ASK with bit-metric decoding.

\section{Conclusion}

\label{sec:conclusion}
We proposed a protograph-based LDPC code design approach for 
bandwidth-efficient coded modulation that is suitable both for uniform 
and shaped inputs. The different bit-channels are replaced 
by biAWGN surrogates so that PEXIT and differential evolution give ensembles with 
good decoding thresholds. The performance of the new codes for uniform inputs are as good as the best codes in literature. For shaped inputs, the new codes operate close to $\frac{1}{2}\log_2(1+\text{SNR})$. Future research should investigate the influence of the surrogates on the code 
performance by employing a full-fledged density evolution for protographs. Furthermore, precoded protographs should be considered to improve the threshold without increasing the variable node degrees.

\newcommand{\BIBdecl}{\setlength{\itemsep}{0.01em}}
\bibliographystyle{IEEEtran}
\normalsize
\bibliography{IEEEabrv,confs-jrnls,literature}

\end{document}